\documentclass[aps,showpacs,superscriptaddress,preprint]{revtex4}
\usepackage{graphicx}
\usepackage{amssymb}
\begin{document}
\title{Axial vector form factor of nucleons in a light-cone
diquark model}
\author{Bo-Qiang Ma}
\email{mabq@phy.pku.edu.cn}
\affiliation{Department of Physics, Peking University, Beijing 100871, China}
\author{Di Qing}
\email{diqing@fis.utfsm.cl}\altaffiliation{corresponding author.}
\author{Iv\'an Schmidt}
\email{ischmidt@fis.utfsm.cl}
\affiliation{Departamento de F\'\i sica, Universidad T\'ecnica Federico
Santa Mar\'\i a, Casilla 110-V, Valpara\'\i so, Chile}

\begin{abstract}
The nucleon axial vector form factor is investigated in a
light-cone quark spectator diquark model, in which Melosh
rotations are applied to both the quark and vector diquark. It is
found that this model gives a very good description of available
experimental data and the results have very little dependence on
the parameters of the model. The relation between the nucleon
axial constant and the anomalous magnetic moment of nucleons is
also discussed.
\end{abstract}

\pacs{13.10.+q, 12.39.Ki, 14.20.Dh}

\maketitle

Since the momenta of quarks within a hadron are of the same order
as the masses of quarks, fully relativistic quark models are
required in order to describe hadron transition processes. The
light-cone quantization provides such a fully relativistic
treatment with lots of unique properties \cite{brodsky98}. It is
well known that the matrix elements of local operators such as the
electromagnetic and weak currents have exact representation in
terms of light-cone wave functions of Fock states
\cite{brodsky01}. If one chooses the special frame \cite{drell70}
$q^+=0$ for the space-like momentum transfer and takes the matrix
elements of plus components of currents, the contribution from
pair creation or annihilation is forbidden and the matrix elements
of space-like currents can be expressed as overlaps of light-cone
wave functions with the same number of Fock constituents.
Therefore, the light-cone frame is well suited for the description
of electromagnetic and weak transition processes.

In previous papers, a light-cone quark spectator diquark model was
proposed in order to investigate the nucleon spin problem
\cite{ma91,ma96,ma98}. This model is based on the assumption that
deep inelastic scattering is well described by the impulse
approximation picture of the quark-parton model \cite{feynman69},
in which the incident lepton scatters incoherently off a quark in
the nucleon, with the remaining nucleon constituents treated as a
quasi-particle spectators to provide the remaining nucleon quantum
number. In fact, in the quark spectator diquark form, some
non-perturbative effects between the two spectator quarks or other
non-perturbative gluon effects in the nucleon can be effectively
taken into account by the mass of the diquark spectator. After
taking into account Melosh rotation effects, this model is in good
agreement with experimental data of polarized deep inelastic
scattering, and the mass difference between the scalar and vector
spectators reproduces the up and down valence quark asymmetry
\cite{ma91,ma96,ma98}.

Recently, based on the impulse approximation, this model was
extended to study the electromagnetic form factors of nucleons and
the results agree with experiment \cite{ma02}. After applying
Melosh rotations to both quark and spectator vector diquark, the
difference between the scalar and vector diquarks breaks the
$SU\left(6\right)$ symmetry of nucleon wave functions and
reproduces the correct electromagnetic nucleon properties,
especially in the case of the neutron. It is natural to extend the
light-cone quark diquark model to study other transition
processes, such as the nucleon axial decay transition process
which is important in the study of the structure of nucleons \cite{Bernard02}.
This is clearly a non-trivial extension, since we are going now into
the weak interactions domain.

The nucleon axial vector form factor $G_A\left(Q^2\right)$ is defined by
\begin{eqnarray}
\left\langle P^\prime, S^\prime\left|A^\mu_a\left(0\right)\right|P,S\right\rangle
=\bar{u}\left(P^\prime, S^\prime\right)\left[G_A\left(Q^2\right)\gamma^\mu+
G_P\left(Q^2\right)\frac{q_\mu}{2M}\right]\gamma_5 \frac{\tau_a}{2}
u\left(P, S\right),\nonumber\\
\end{eqnarray}
where $A^\mu_a\left(0\right)$ is the axial vector current,
$q^\mu=\left(P^\prime-P\right)^\mu$ is four-momentum transfer,
$Q^2=-q^2$, $u\left(P, S\right)$ is the nucleon spinor, and
$\tau_a$ is the isospin matrix with Cartesian index $a$. In the
light-cone frame, the plus component of the axial vector current
reads
\begin{equation}
\frac{\bar{u}\left(k^\prime,\uparrow\right)}{\sqrt{k^{\prime +}}}
\gamma^+\gamma_5 \frac{u\left(k,\uparrow\right)}{\sqrt{k^+}}
=-\frac{\bar{u}\left(k^\prime,\downarrow\right)}{\sqrt{k^{\prime +}}}
\gamma^+\gamma_5 \frac{u\left(k,\downarrow\right)}{\sqrt{k^+}}=2,
\end{equation}
and in these calculations we choose the Drell-Yan assignment \cite{drell70}:
\begin{eqnarray}
q&=&\left(q^+,q^-,\vec{q}_{\perp}\right) = \left(0,\frac{-q^2}{P^+},
\vec{q}_{\perp}\right), \nonumber \\
P&=&\left(P^+,P^-,\vec{P}_{\perp}\right) = \left(P^+,\frac{M^2}{P^+},
\vec{0}_{\perp}\right),
\end{eqnarray}
thus we have
\begin{equation}
G_A\left(Q^2\right)=\left\langle P^\prime,\uparrow\left|
\frac{A^+\left(0\right)}{2P^+}\right|P,\uparrow\right\rangle.
\end{equation}
Thus, in similar to the electromagnetic operators, the matrix
element of axial vector currents can be expressed in the
light-cone formalism as overlaps of light-cone wave functions with
the same number of Fock constituents as:
\begin{equation}\label{eqn:overlap}
G_A\left(Q^2\right)=\sum_a\int\frac{d^2\vec{k}_\perp dx}{16\pi^3}
\sum_j \tau_{j}\lambda_j\psi^{\uparrow\star}_a\left(x_i,\vec{k}^\prime_{\perp i},
\lambda_i\right) \psi^{\uparrow}_a\left(x_i,\vec{k}_{\perp i},
\lambda_i\right),
\end{equation}
where $\tau_{j}$ and $\lambda_j$ are the isospin and helicity of the struck
constituents, $\psi^{\uparrow}_a
\left(x_i,\vec{k}^\prime_{\perp i},\lambda_i\right)$ is the light-cone
Fock expansion wave function, and $\lambda_i$, $x_i$ and $\vec{k}_{\perp i}$
are the spin projections along the quantization $z$ direction, light-cone
momentum fractions and relative momentum coordinates of
QCD constituents, respectively. Here, for the final state light-cone
wave function, the relative momentum coordinates are
\begin{equation}
\vec{k}^\prime_{\perp i}=\vec{k}_{\perp i}+\left(1-x_i\right)\vec{q}_{\perp}
\end{equation}
for the struck quark and
\begin{equation}
\vec{k}^\prime_{\perp i}=\vec{k}_{\perp i}-x_i\vec{q}_{\perp}
\end{equation}
for each spectator.

In this work we study the nucleon axial vector form factor
based on the light-cone quark spectral diquark model \cite{ma02},
in which the Melosh rotations is applied to both quark and diquark,
explicitly
\begin{eqnarray}\label{eqn:melosh}
\chi^{\uparrow}_T &=& w \left[\left(k^+ + m\right)
\chi^{\uparrow}_F - k^R \chi^{\downarrow}_F\right],
\nonumber \\
\chi^{\downarrow}_T &=& w \left[\left(k^+ + m\right)
\chi^{\downarrow}_F - k^L \chi^{\uparrow}_F\right],
\end{eqnarray}
for quarks \cite{melosh74}, and
\begin{eqnarray}\label{eqn:melosh1}
V^1_T &=& w^2 \left[\left(k^+ +m\right)^2 V^1_F
-\sqrt{2}\left(k^+ +m\right)k^R V^0_F + {k^R}^2 V^{-1}_F\right],\nonumber\\
V^0_T &=& w^2 \left[\sqrt{2}\left(k^+ +m\right)k^L V^1_F
+ 2\left(\left(k^0+m\right)k^+-k^R k^L\right)V^0_F \right.\nonumber\\
 &&\left. -\sqrt{2}\left(k^+
+m\right)k^R V^{-1}_F\right], \nonumber\\
V^{-1}_T &=& w^2 \left[{k^L}^2 V^1_F+\sqrt{2}\left(k^+ +m\right)k^L
V^0_F + \left(k^+ +m\right)^2V^{-1}_F\right],
\end{eqnarray}
for vector diquarks \cite{ahluwalia93}. Here, $\chi_T$ and $\chi_F$
are instant and light-cone spin-$\frac{1}{2}$ spinors, $V_T$ and $V_F$
are the instant and light-cone spin-$1$ spinors respectively,
$w=\left[2k^+\left(k^0+m_q\right)\right]^{-\frac{1}{2}}$,
$k^{R,L}=k^1\pm i k^2$, and $k^+ = k^0+ k^3$.
And the details of the quark diquark model can be found in Ref. \cite{ma02}.
Therefore, according to Eq. (\ref{eqn:overlap}), we have
\begin{eqnarray}\label{eq:fm}
G_A\left(Q^2\right)&=&3\int\frac{d^2 k_\perp dx}{16\pi^3}w^\prime_q w_q
\left\{\frac{1}{9} \sin^2\theta \left\{-\left[\left(k^{\prime +}_q+
m_q\right)\left(k^+_q+m_q\right)-k^{\prime L}_\perp k^R_\perp\right]
\right.\right.\nonumber\\
&&\left.\left.\times O_{V^{0,0}}
+\sqrt{2}\left[\left(k^{\prime +}_q+m_q\right)k^L_\perp +
\left(k^+_q+m_q\right)k^{\prime L}_\perp\right]O_{V^{0,1}}
\right.\right.\nonumber \\
&& \left.\left.+\sqrt{2}\left[\left(k^{\prime +}_q+m_q\right)k^R_\perp +
\left(k^+_q+m_q\right)k^{\prime R}_\perp\right]O_{V^{1,0}}
\right.\right.\nonumber \\
&& \left.\left.+2\left[\left(k^{\prime +}_q+
m_q\right)\left(k^+_q+m_q\right)-k^{\prime R}_\perp k^L_\perp\right]
O_{V^{1,1}}\right\}\varphi_V\left(x,\vec{k}^\prime_\perp\right)
\right.\nonumber\\
&& \left. \times\varphi_V\left(x,\vec{k}_\perp\right)
 +\cos^2\theta\left[\left(k^{\prime +}_q+
m_q\right)\left(k^+_q+m_q\right)-k^{\prime L}_\perp k^R_\perp\right]
\right.\nonumber\\
&& \left. \times \varphi_S\left(x,\vec{k}^\prime_\perp\right)
\varphi_S\left(x,\vec{k}_\perp\right)\right\},
\end{eqnarray}
where $O_V$ come from the Melosh rotation of vector diquarks, and $\varphi$
are the momentum space wave function which is assumed to be a harmonic
oscillator wave function (the Brodsky-Huang-Lepage (BHL) prescription
\cite{brodsky81}).

\begin{figure}[b]
\includegraphics{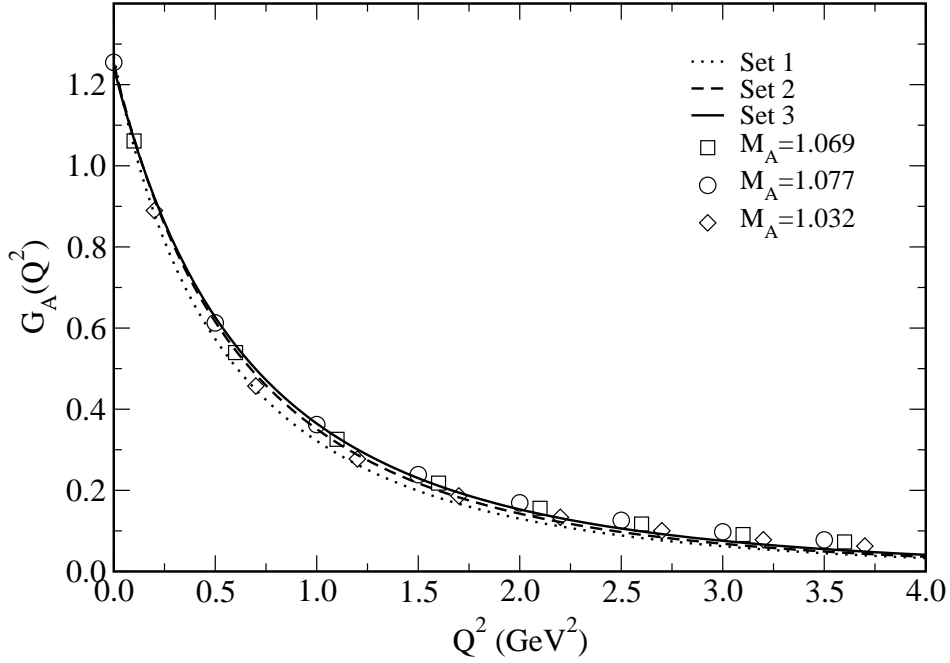}
\caption{\label{fig:axial}Axial form factor of nucleons. The experimental
data are from \cite{liesenfeld99,kitagaki83}}
\end{figure}

Following Ref. \cite{ma02}, we calculate the nucleon axial vector form
factor using three different sets of parameters in order to show its
dependence on the difference between the scalar and vector diquarks.
This is shown in Fig. \ref{fig:axial}. As Ref. \cite{glozman01}, the
experimental data are assuming to be the dipole form
\begin{equation}
G_A\left(Q^2\right)=\frac{g_A}{\left(1+Q^2/M_A^2\right)^2},
\end{equation}
where the axial constant $g_A=1.2670(35)$ is from the most recent
review by Particle Data Group \cite{pdg00}, and $M_A$ is the axial
mass. Our results agree with the experiment very well. In contrast
to the case of nucleon electromagnetic form factors, the nucleon
axial vector form factor is largely parameter independent.
Alternatively, the effect of the difference between the scalar and
vector diquarks to the the nucleon axial vector form factor is
small. The static properties corresponding to the axial vector
current, axial constant $g_A$ and axial radius $\left\langle r^2_A
\right\rangle^{1/2}$, are listed in Table \ref{tab:properties}.
The axial radius is calculated by
\begin{equation}
r^2_A = \left. -6\frac{1}{G_A\left(0\right)}\frac{d G_A\left(Q^2\right)}{d Q^2}
\right|_{Q^2=0}.
\end{equation}
Our results are very close to the experimental data, and also the
effect of the difference between the scalar and vector diquarks to
the the nucleon axial vector static properties is small.

For the axial constant $g_A=G_A\left(0\right)$, the relativistic
effects are essential in order to reduce the $SU\left(6\right)$ nonrelativistic
result $g^{NR}_A=5/3$ to the experimental data. In fact, from Eq. (\ref{eq:fm})
we have
\begin{equation}
g_A=\int\frac{d^2 k_\perp dx}{16\pi^3}\left[ \frac{1}{6}
W_A\left(x,\vec{k}_\perp\right)\varphi^2_V\left(x,\vec{k}_\perp\right)
+\frac{3}{2}W_A\left(x,\vec{k}_\perp\right)
\varphi^2_S\left(x,\vec{k}_\perp\right)\right],
\end{equation}
where
\begin{equation}
W_A\left(x,\vec{k}_\perp\right)=\frac{\left(k^+_q+m_q\right)^2-
\vec{k}^2_\perp}{\left(k^+_q+m_q\right)^2+\vec{k}^2_\perp},
\end{equation}
is the Wigner rotation factor corresponding to the contribution
from the relativistic effects due to the quark transversal
motion \cite{ma91,ma96,chung91}. In the case of SU(6) symmetry
between the vector and scalar diquarks, we get the same results as
other light-cone quark models \cite{chung91,brodsky94},
$g_A=\left\langle W_A\right\rangle g^{NR}_A$. In the
nonrelativistic limit, $\left\langle W_A\right\rangle = 1$ and
$g_A=g^{NR}_A$. Therefore, the physical value of the axial
constant is reduced $25\%$ from its nonrelativistic value due to
relativistic effects. This is similar to the nucleon spin problem
situation \cite{ma91}, where relativistic effects are
important for producing the quark spin reduction on the
light-cone. This is reasonable, since from Eq. (\ref{eqn:overlap})
it is easy to obtain the following relation between the nucleon
axial constant and the quark spin contributions to the nucleon
spin,
\begin{equation}
g_A=\Delta u -\Delta d.
\end{equation}
Here $\Delta u$ and $\Delta d$ are the helicity of up and down quarks
in the nucleon, and $\Delta q= \left\langle W_A\right\rangle \Delta
q^{NR}$, where $\Delta q^{NR}$ is the nonrelativistic quark spin
contributions to the nucleon spin defined in the quark model.

On the other side, in our model the anomalous magnetic moment of proton
$a$ can be also written as \cite{ma02}
\begin{eqnarray}
a&=&2M\int\frac{d^2 k_\perp dx}{16\pi^3}\frac{1}{\mathcal{M}}\left[
\frac{(1-x)\mathcal{M}\left(k^+_q+m_q\right)-\vec{k}^2_\perp/2}
{\left(k^+_q+m_q\right)+\vec{k}^2_\perp}\right]
\varphi_S^2\left(x,\vec{k}_\perp\right) \nonumber\\
&=&\left \langle W_M \right \rangle a^{NR},
\end{eqnarray}
where $a^{NR}=2M/3m_q$ is the nonrelativistic value of the proton
anomalous magnetic moment, and the relativistic effect corresponding to
the anomalous magnetic moment
\begin{equation}
W_M=\frac{3m_q}{\mathcal M}\left[\frac{(1-x)\mathcal{M}\left(k^+_q+m_q\right)
-\vec{k}^2_\perp/2}{\left(k^+_q+m_q\right)+\vec{k}^2_\perp}\right],
\end{equation}
which is the same as Ref. \cite{brodsky94}. Thus, the
wavefunction-independent
relations between the nucleon axial-coupling $g_A$ and the
nucleon magnetic moments of Ref. \cite{brodsky94} are still kept
in our model.

In conclusion, we extended our studies of nucleon elastic and
inelastic scattering processes in a light-cone quark
diquark model to the nucleon axial vector form factor
and the Melosh rotations were applied to both the quark and vector
diquark. It is shown that the axial vector form factor has very little
dependence on the parameters of the model and
the relativistic properties of the model
are essential for the good agreement with the experimental
results. The relation between the nucleon axial constant and the
anomalous magnetic moment of nucleons is also discussed.

Therefore the light-cone quark diquark model gives a very good
description of the nucleon. In fact, we have shown that it can
reproduce quite accurately both elastic and inelastic
electromagnetic and weak nucleon data, as can be seen by our
results for the nucleon structure functions and vector and axial
vector form factors.

\acknowledgments{This work is partially supported by National
Natural Science Foundation of China under Grant Numbers 19975052,
10025523, and 90103007, by Fondecyt (Chile) under project 3000055
and Grant Number 8000017.}

\newpage

\begin{table}
\caption{\label{tab:properties}The static electromagnetic properties
of nucleons for the three sets of parameters.}
\begin{ruledtabular}
\begin{tabular}{lrrrr}
        & Set I & Set II & Set III & Expt. \\
\hline
$g_A$ & 1.253 & 1.270 & 1.242 & 1.2670(35)\cite{pdg00} \\
$\left\langle r^2_A\right\rangle^{1/2}$ & 0.624 & 0.703 & 0.611
& 0.635(23)\cite{liesenfeld99},0.65(7)\cite{kitagaki83} \\
\end{tabular}
\end{ruledtabular}
\end{table}

\end{document}